\documentstyle[11pt,newpasp,twoside]{article}

\include{psfig} \psnoisy 

\begin{document}

\title{The Spectral Appearance of Primeval Galaxies}
\author{Bruno Guiderdoni \& Julien E.G. Devriendt}
\affil{Institut d'Astrophysique de Paris, CNRS, 98bis Bld Arago, 75014 Paris}

\begin{abstract}
The current and forthcoming observations of large samples of high--redshift 
galaxies selected according to various photometric and spectroscopic criteria 
can be interpreted in the context of galaxy formation, by means of models 
of evolving spectral energy distributions (SEDs). We hereafter present 
{\sc stardust} which gives synthetic SEDs from the far UV to the 
submm wavelength range. These SEDs are 
designed to be implemented into semi--analytic models of galaxy formation.
\end{abstract}

\keywords{Galaxies : spectra, Galaxies : formation, Galaxies : evolution}

\section{The zoo of high--redshift galaxies}
The search for primeval galaxies has been one of the long--term programs of 
observational cosmology, along with the development of sensitive detectors, 
and is currently receiving an exciting boost with the new generation of 
8--metre class telescopes. However, the search itself is somewhat hampered by 
the fuzziness of the concept of ``primeval galaxy''. This is generally
understood as being ``a galaxy which is captured at the epoch of its 
formation''. Since there are several competing theories about what a 
forming galaxy should look like, it is readily possible that we are missing 
part of the 
process of galaxy formation because we do not know yet what to search for. 

More specifically, two general paradigms have been proposed in the last twenty 
years or so. In the picture of monolithic collapse, galaxies form at a given 
epoch $z_{for}$, when the physical conditions of the universe are favourable, 
and evolve at different rates which are fixed by the 
initial conditions. In the picture of hierarchical collapse, there is nothing 
like a given redshift of galaxy formation. Larger galaxies form from the 
merging of smaller ones, which on their turn have formed from the merging of 
still smaller lumps, and so on. The beginning of the process took place
at some early redshift $z \sim 30$ (when the first objects can cool) 
and is still going on now. As a result, the ``epoch of galaxy formation'' can
be defined e.g. as the epoch when the first stars formed, when 50 \% of the 
stars have formed, or when the 
morphology was fixed after the last major merging event. This hierarchical
galaxy formation is now modelled in the context of hierarchical clustering 
where dark matter completely rules gravitational collapse.

The models of galaxy formation have to reproduce the wide variety of objects 
which are now observed at high redshift, probably after strong observational 
biasses that are not fully understood. These objects are generally selected 
according to photometric criteria. We now have Luminous Blue Compact Galaxies
(LBCGs; $z\sim 1$ and $I_{AB} < 22.5$),
Lyman--Break Galaxies (LBGs; $z\simeq 3$ or $z\simeq 4$, and $I_{AB} < 25$), 
Extremely Red Objects (EROs; $R-K >6$ and $K < 19.5$), 
Lyman Alpha Galaxies (LAGs; $z\simeq 4$ or $z\simeq 5$, and $EW({\rm Ly}
\alpha) > 100$ \AA), faint submm sources
that are likely to be the high--$z$ counterparts of the local ``Ultraluminous 
Infrared Galaxies'' discovered by {\sc iras} (high--$z$ ULIRGs;
$z>1$ and $L_{IR} > 10^{12}$ $L_\odot$), Damped Lyman $\alpha$ 
Absorbers (DLAs; $N_H > 10^{22}$ atom cm$^{-2}$), etc. 
In order to link the predictions of the models of galaxy formation with these
observations, it is necessary to model the spectral energy distributions (SEDs)
on the widest wavelength range, from the rest--frame UV to the rest--frame 
submm.

\begin{figure}[h]
\centerline{
\psfig{figure=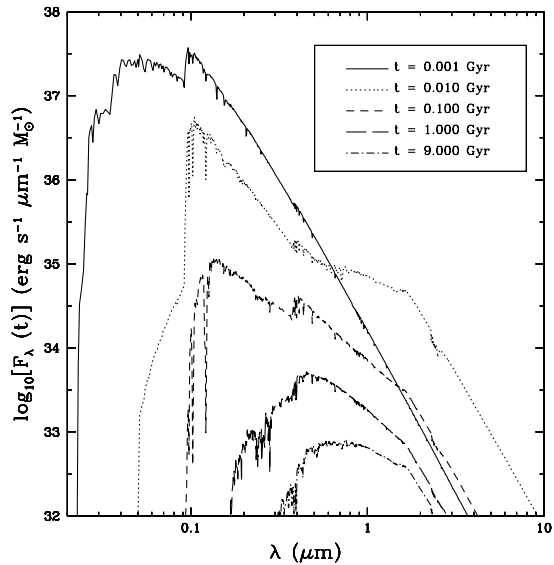,width=0.6\textwidth}}
\caption{\small Spectral evolution of an instantaneous burst of star 
formation with solar metallicity. The IMF is Salpeter and the flux level 
is reduced to $1~M_\odot$ of galaxy.}
\end{figure}

\section{The case for high--redshift dust}
As a matter of fact, several pieces of observational evidence are now 
converging to show that there is a significant amount of extinction in 
high--redshift galaxies. Consequently, the absorption and emission processes 
due to dust cannot be neglected in assessing the luminosity budget of forming 
galaxies. Among these pieces of evidence, we can quote:
 
(i) The discovery of the Cosmic IR Background at a level 10 times 
higher than the no--evolution predictions based on the {\sc iras} luminosity 
functions, and twice as high as the Cosmic Optical Background (Puget
{\it et al.} 1996, Guiderdoni {\it et al.} 1997, Fixsen {\it et al.} 1998,
Hauser {\it et al.} 1998, Lagache {\it et al.} 1999).

(ii)  The IR and submm counts that have broken the CIRB into its brightest 
contributors at 15 $\mu$m (ISOCAM down to $\sim 0.1$ mJy, Aussel {\it et al.}
1999, Elbaz {\it et al.} 1999), 175 $\mu$m 
(ISOPHOT down to $\sim 100$ mJy, Kawara {\it et al.} 1998, 
Puget {\it et al.} 1999), and 850 $\mu$m (SCUBA down to 2 mJy, Smail 
{\it et al.} 1997, Hughes {\it et al.} 1998, Barger {\it et al.} 1998, 1999a, 
Eales {\it et al.} 1999).
Although the poor spatial resolution of the observations makes the 
identification of the optical counterparts somewhat difficult, the first 
results of the spectroscopic follow--ups seem to show that some of the 
sources are at $z>1$ (Smail {\it et al.} 1998, Lilly {\it et al.} 1999, 
Barger {\it et al.} 1999b).

(iii) At least some of the EROs (e.g. HR10 at $z=1.44$, Cimatti {\it et al.} 
1998a) are high--redshift dusty objects.

\begin{figure}[h]
\centerline{
\psfig{figure=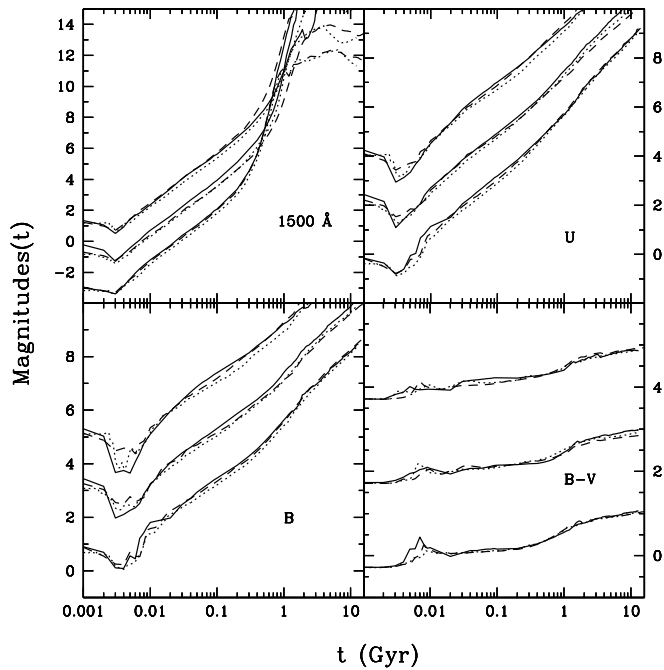,width=0.7\textwidth}}
\caption{\small Comparison of {\sc gissel} 1998 (dotted line),
{\sc p\'egase} (dashed line), and {\sc stardust} (solid line) for an 
instantaneous burst with solar metallicity (lowest curve), 2/5 solar 
(middle curve), and 1/5 solar (upper curve) in the different wavelength bands 
indicated on the panels. For a better visualization, the 
1/5 solar and 2/5 solar metallicities have been
shifted by 2 and 4 magnitudes respectively.}
\end{figure}

(iv) The extinction in LBCGs at $z\sim 1$ corresponds to a factor 3 
extinction on the rest--frame flux at 2800 \AA\ (Flores {\it et al.} 1999).

(v) The extinction in LBGs at $z\sim 3$ and 4 is high ($0.1 \le E(B-V) \le
0.5$), with a trend of a larger extinction for the brightest objects
(Steidel {\it et al.} 1999, Meurer {\it et al.} 1999). This 
corresponds to a factor 5 extinction on the rest--frame flux at 1600 \AA.

(vi) Dust is present in high--redshift radiogalaxies and QSOs, up to $z=4.69$
(see e.g. Hughes {\it et al.} 1997, Cimatti {\it et al.} 1998b).
The optical spectrum of a gravitationally--lensed galaxy at $z=4.92$ 
already shows a reddening factor amounting 
to $0.1 < E(B-V) < 0.3$ (Soifer {\it et al.} 1998).

(vii) The correlation of dust with metallicity has still to be understood, 
since, for instance, the extremely metal--poor galaxy SBS0335-052 
(1/30 of solar) has a significant IR emission (Thuan {\it et al.} 1999).

\section{Spectral energy distributions of young galaxies}
We hereafter present a model of synthetic SEDs, called {\sc stardust},
which contains up--to--date spectrophotometric modelling
coupled to chemical evolution and dust absorption. 
The model allows one to make consistent predictions 
for the SEDs of galaxies in a very broad wavelength range. Details can be 
found in Devriendt {\it et al.} (1999; hereafter DGS).

\begin{figure}[h]
\hbox{\hskip -0.3truecm
\psfig{figure=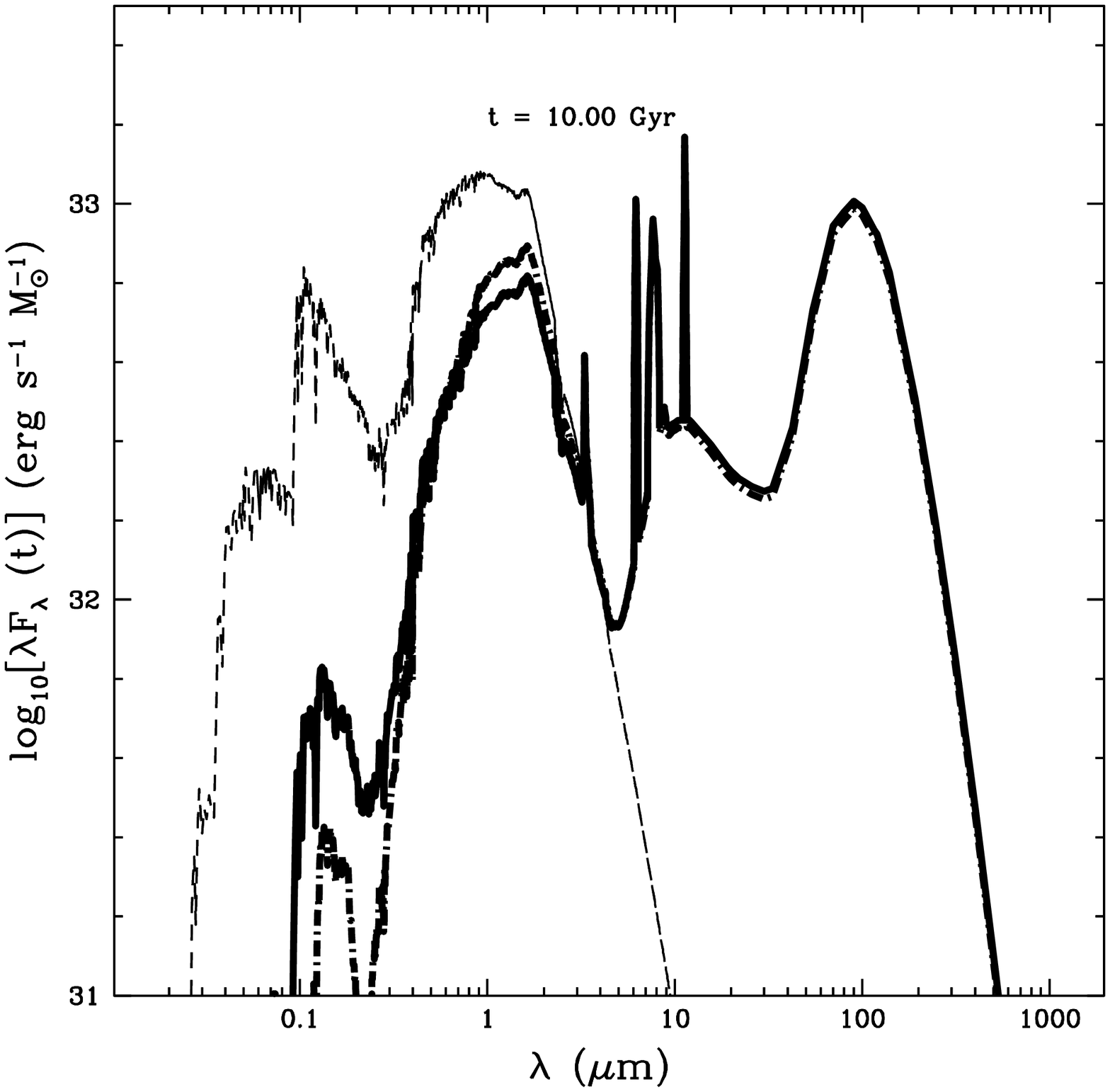,width=0.5\textwidth}
\psfig{figure=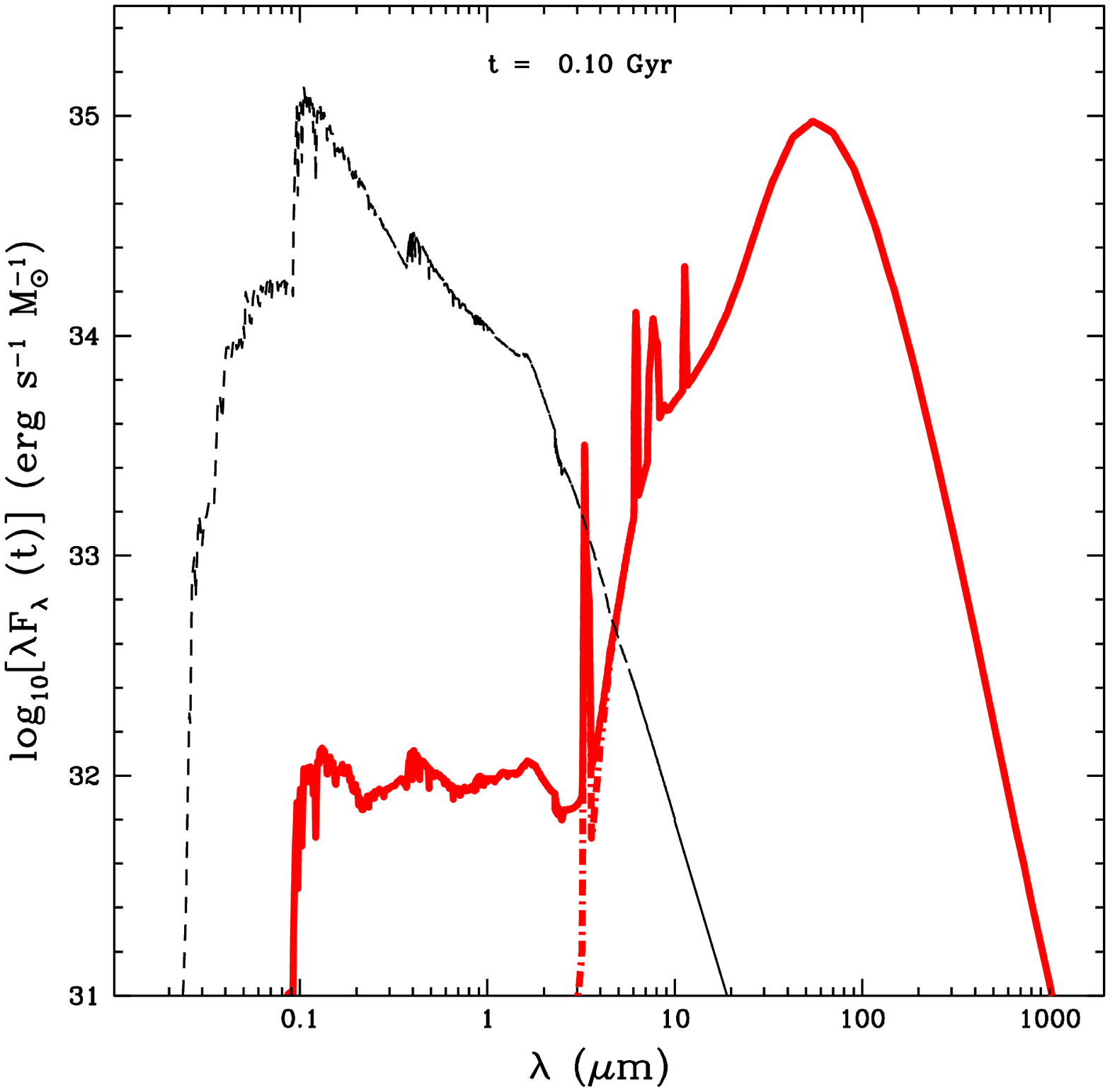,width=0.5\textwidth}
}
\caption{\small {\it Left--hand panel:} Snapshot of the full 
wavelength range synthetic
spectrum of a typical spiral galaxy with a star
formation time scale $t_*=3$ Gyr taken at time $t = 10$ Gyr, 
and $f_H = 1$ corresponding to $\tau_V \simeq 1$. 
The different curves represent different 
geometries of dust and stars. The thick solid line is the homogeneous oblate 
ellipsoid mix, and the thick dot--dashed line assumes a screen geometry. As a
guideline, we also plot (thin dashed curve) the spectrum without any
absorption.  {\it Right--hand panel:}
Typical spectrum of a forming galaxy at time $t = 0.1$ Gyr. The thin line is 
a dust--free primeval galaxy. The thick lines 
correspond to the {\it same} galaxy, but with strong extinction 
($f_H = 100$, that is, $\tau_V \simeq 100$) leading to an ULIRG. 
For the screen geometry (dot--dashed line), 
the level of the flux below 2 $\mu$m is negligible.}
\end{figure}

The model uses the so--called ``isochrone scheme'' and includes a compilation 
of the Geneva tracks (see e.g. Charbonnel {\it et al.} 1996, and references 
therein) for various metallicities $Z$, and masses 
0.8 $\le M/M_{\odot} \le$ 120. For stars less massive
than $1.7 M_{\odot}$, some of the old tracks stop at the Giant Branch tip. 
More recent grids of models, based on Geneva tracks and 
covering the evolution of low mass stars 
(0.8 to 1.7 $M_{\odot}$) from the Zero--Age Main Sequence 
up to the end of the Early--AGB, are included for $Z=$ 0.001 
and $Z=$ 0.02.
For the late stages of other metallicities (Horizontal Branch,
Early--AGB), we either interpolate or 
extrapolate $\log L_{bol}$, $\log T_{eff}$ 
and $\log t$ versus $\log Z$ from the available tracks.  
The final stages of the stellar evolution (Thermally--Pulsing--AGB and 
Post--AGB) are not included. 
We use the grid of theoretical fluxes from Kurucz (1992)
which covers all metallicities 
from $\log Z/Z_\odot = +1.0 $ to $\log Z/Z_\odot =-5.0 $, and 61 
temperatures from 3500 K to 
50 000 K. Each spectrum spans a wavelength range between 
90 $\rm \AA$ and 160 $\mu$m, with a mean resolution of 
20 $\rm \AA$. 
For the coldest stars (K and M-type stars) with $T \le$ 3750 K,
Kurucz's models fail to reproduce the observed spectra.
Therefore, we prefer to use models coming from 
different sources (see DGS). 

Fig 1 gives the passive evolution of a Simple Stellar 
Population with Salpeter IMF (slope $x=1.35$
from $m=0.1$ to $120$ $M_\odot$). Fig 2
compares our {\sc stardust} model with other models available in the 
literature (see DGS for details) : {\sc gissel} 1998 
(Bruzual and Charlot, 1993), {\sc p\'egase} 
(Fioc and Rocca--Volmerange, 1997), and our {\sc stardust}. 
Though the stellar data are different, the agreement is 
generally very good in the UV/visible, except when Post---AGB are dominant
(that is, in old stellar populations without star formation activity).
In particular, the UV to SFR ratio that is used to derive the cosmic SFR from
rest--frame 1600 \AA\ or 2800 \AA\ observations is remarkably similar
in the three models.

\begin{figure}[h]
\centerline{
\psfig{figure=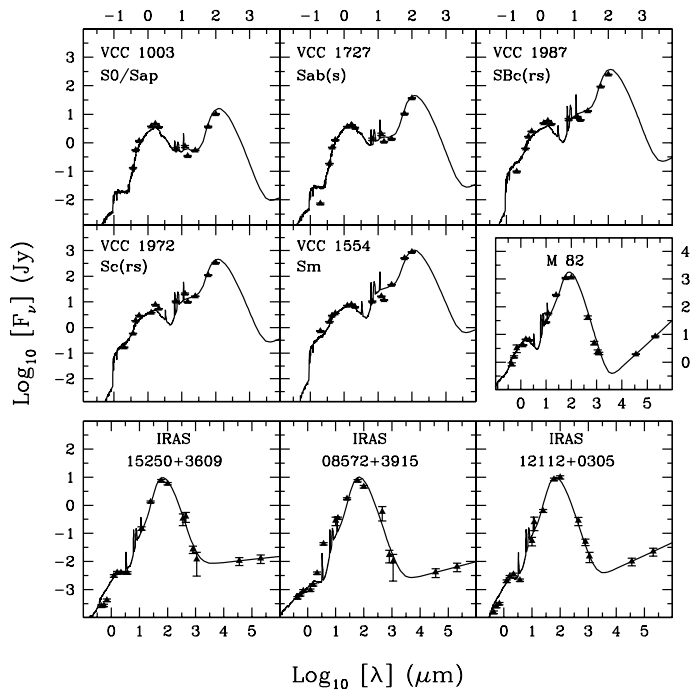,width=0.7\textwidth}}
\caption{\small Best--fit model for galaxies extracted from a sub-sample of 
local spirals and ULIRGs. Objects are ordered with increasing 
$L_{IR}$ from top to bottom and from left to right.}
\end{figure}

Our model computes chemical evolution and takes into account the effect
of metallicity on the stellar tracks and stellar spectra in a consistent way. 
In addition, the metallicity of the gas is followed. Under simple 
assumptions on the variation of the extinction curve with metallicity 
(based on the study of the Milky Way, The Large Magellanic Cloud, and the 
Small Magellanic Cloud), and on the geometry and relative distributions of 
dust and 
stars (homogeneous mix in an oblate spheroid), transfer can be easily solved, 
and the amount of luminosity 
absorbed by dust is estimated. The SEDs in the IR/submm are then computed 
to reproduce the correlation of {\sc iras} colours with total IR luminosity
$L_{IR}$, following Guiderdoni {\it et al.} (1997, 1998).
The model reproduces the obscuration curve of Calzetti 
{\it et al.} (1994) observed in a sample of nearby starbursts. 
It also reproduces the 
correlation of IR to 2200 \AA\ flux with the UV slope $\beta$ of the SEDs 
(around 2200 \AA) observed by Meurer {\it et al.} (1995). 

\begin{figure}[h]
\centerline{
\psfig{figure=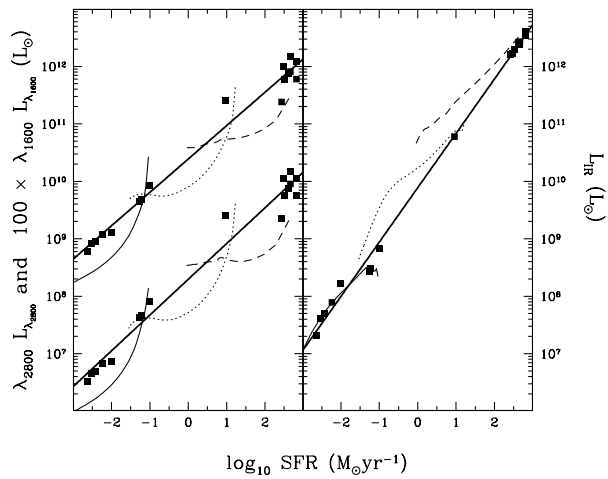,width=0.6\textwidth}}
\caption{\small Star formation rate 
against luminosity at 2800 \AA \, and 1600 \AA \, (left panel), and 
total infrared 
luminosity (right panel). Symbols represent quantities derived from
our best fit models. For clarity, both these quantities and curves at 
1600 \AA \,
have been arbitrarily shifted up by two decades. The evolution for
three models is plotted with solid lines, dotted lines, and dashes
(see DGS). The curves show time evolution 
for ages ranging from $t=$ 0.01 (maximum star formation rate and
luminosities) to 15 Gyr (minimum star formation rate and
luminosities). The straight lines represent least square fits.} 
\end{figure}

Finally, the optical and IR/submm spectra are connected to give evolving 
synthetic SEDs from the far UV (90 \AA) to the submm ($\sim 1$ mm). In the 
centimetre and metre range, a radio component can be added under 
assumptions on the slope and on the correlation of radio fluxes with IR fluxes.
The SEDs depend on three parameters (in addition to the IMF) : the SFR 
timescale $t_*$, the age $t$, and a concentration parameter $f_H$ that 
describes the size of the gaseous disk ($f_H=1$ is for normal spirals, and
a radial collapse by a factor 10 corresponds to $f_H=100$). Fig 3 gives
typical spectra for a spiral and a forming galaxy, without and with extinction
(respectively with $f_H=1$ and $f_H=100$, this latter value being typical of 
an ULIRG). 

\section{From local to high--redshift ULIRGs}
Fig 4 gives the fits of a sample of local galaxies obtained with our 
theoretical SEDs and a $\chi^2$ procedure. The sample gathers Virgo Cluster 
spirals (Boselli {\it et al.} 1998) and ULIRGs (Rigopoulou {\it et al.} 1996) 
that 
have a sufficient number of photometric points at optical, NIR, MIR, FIR, and 
submm wavelengths (and, for some of them, in the radio). Nine spectra have 
been ordered in fig 4 according to 
their $L_{IR}$, in a sequence that parallels the compilation of Sanders 
\& Mirabel 
(1996). There is some degeneracy between the parameters 
$t_*$ and $t$ of the fits, but their combination corresponds to similar SFRs. 
This is illustrative of the ability of the model to
capture the characteristic features of the objects.

This can be used to study the efficiency of the UV and IR fluxes to trace 
the underlying SFR. As shown in fig 5, the UV fluxes do scale with SFR for 
optically--thin galaxies (slope $\simeq 1$), but the proportionality breaks 
down for ULIRGs. Without any 
information on the optical thickness of the high--redshift galaxies, 
the rest--frame UV fluxes could lead to erroneous estimates. 
In contrast, the IR fluxes trace SFRs on several orders 
of magnitudes, with a slope 0.95, and independently of the optical thickness. 
The origin of this behaviour is not clear yet. It is probably due to the fact 
that there are basically two regimes : either the galaxy is optically thin, 
and the IR roughly corresponds to a constant fraction of the UV, or it is 
optically thick, and all the UV flux is absorbed by dust and released in 
the IR. 

\begin{figure}[htbp]
\centerline{
\psfig{figure=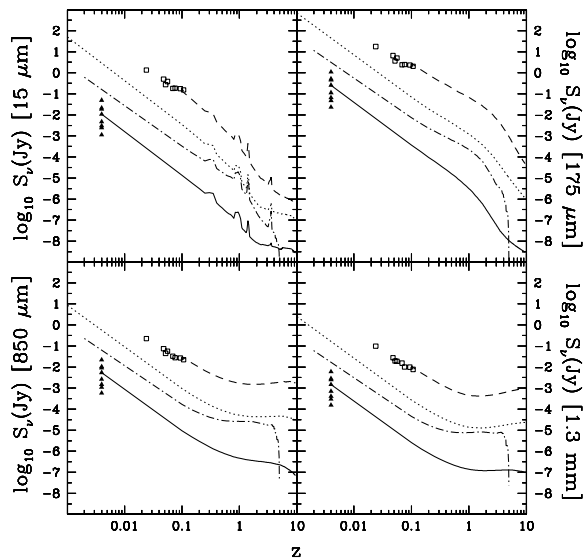,width=0.6\textwidth}}
\caption{\small Apparent IR/submm fluxes for 3 objects,
the ``normal'' spiral VCC 836 of the Virgo Cluster (solid line), the
close--by ``mild starburst'' M82 (dotted line),
and the ULIRG IRAS 14348-1447 (dashed line), as a function of
redshift, for a flat cosmology
where $H_0 =$ 65 km s$^{-1}$ Mpc$^{-1}$, $\Omega_0 = 0.3$ and
$\Omega_\Lambda = 0.7$. The symbols are galaxies of our sample. The
dot--dashed line is an $M = 10^{10}~M_\odot$
model spiral with $t_* = 3$ Gyr that formed at redshift
5, and for which the evolution correction is included.}
\end{figure}

These galaxies can also be used to predict the photometric properties at 
higher redshift. Fig 6 and 7 give the observer--frame optical magnitudes 
and IR/submm fluxes versus redshift.  

\begin{figure}[htbp]
\centerline{
\psfig{figure=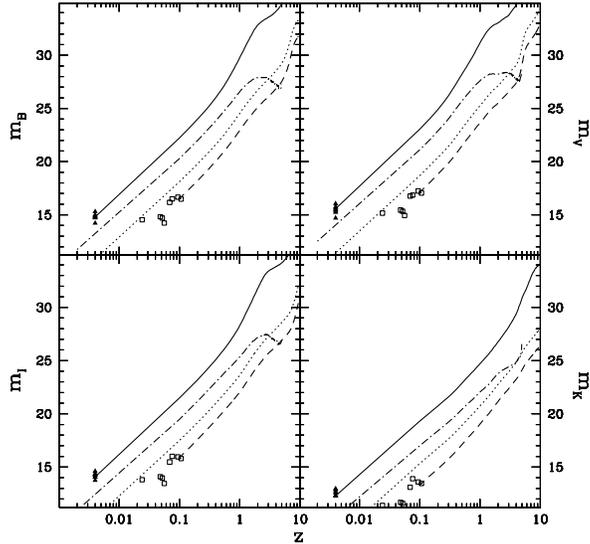,width=0.6\textwidth}}
\caption{\small Apparent magnitudes for the same objects in  
different optical and NIR wavebands, as a function of redshift. 
Coding for the lines is the same as in the previous figure.}
\end{figure}

\section{Semi--analytic and hybrid models of galaxy formation}
These synthetic SEDs can subsequently be implemented into semi--analytic models
of galaxy formation. The interest of such an approach is that
the distribution of the SFR timescales $t_*$ can be 
computed from the distribution of dynamical timescales $t_{dyn}$ of the host
haloes, under the assumption $t_*=\beta t_{dyn}$, where $\beta$ is 
an efficiency parameter which is fixed observationally 
(see e.g. Kennicutt 1998). The ages $t$ of the stellar populations are 
computed from the formation redshifts and the redshifts 
at which the galaxies are observed. 
The distribution of the gas column densities is also computed from the
size distribution of disks (obtained from the size distribution of haloes 
after conservation of angular momentum). The ULIRGs correspond to a further 
radial collapse by a factor 10 ($f_H=100$). 

Following Guiderdoni {\it et al.} (1997, 1998), Devriendt \& Guiderdoni (1999) 
proposed results of faint galaxy counts that are produced by a simple,
semi--analytic prescription using the peaks formalism. The usual recipes are 
implemented for gas cooling, dissipative collapse, and stellar feedback. 
It is assumed that dust is heated by starbursts, and that the Star 
Formation Rate is $SFR(t)=M_{gas}/t_*$, with $t_* \equiv \beta t_{dyn}$. The
spectra are taken from {\sc stardust}. This paper gives predictions of faint 
counts at optical, IR, and submm wavelengths, for a variety of cosmologies.
It turns out that the ionizing flux escaping from high--redshift 
galaxies estimated with this
model is unable to ionize the intergalactic medium (Devriendt {\it et al.}
1998).

\section{Conclusions and prospects}
These synthetic SEDs and the spectral templates of fig 4 are useful tools to 
analyse the panchromatic view on high--redshift, young and ``primeval'' 
galaxies. The spectral templates are available upon request to the authors. 
The implementation of such spectra into semi--analytic models is a promising 
way to predict the statistical properties of a host of objects
that are now observed at high redshift, through various spectral windows, and
with various selection criteria.

However, the modelling of the SFR history of galaxies in the context of 
hierarchical galaxy formation is still incomplete. In particular, the 
ULIRGs appear to be key objects to understand the formation of bulges. 
A more realistic treatment of such objects requires the monitoring of the
merging rates, and involves keeping track of the spatial and dynamical 
information in models of galaxy formation. This can be done, for instance, 
with the new generation of the so--called ``hybrid'' models in which the
merging history trees are directly built from the outputs of cosmological 
N--body simulations. The implementation of {\sc stardust} into such models
is in progress.


\begin{references}
\reference Aussel, H., Cesarsky, C.J., Elbaz, D., Starck, J.L., 1999,
{\it A\&A}, 342, 313
\reference Barger, A.J., Cowie, L.L., Sanders, D.B., Fulton, E., 
Taniguchi, Y., Sato, Y., Kawara, K., Okuda, H., 1998, {\it Nature}, 394, 248
\reference Barger, A.J., Cowie, L.L., Sanders, D.B., 1999a, {\it ApJ}, 518, L5
\reference Barger, A.J., Cowie, L.L., Smail, I., Ivison, R.J., Blain, W., 
Kneib, J.P., 1999b, {\it AJ}, 117, 2656
\reference Boselli, A., Lequeux, J., Sauvage, M., Boulade, O., Boulanger, 
F., Cesarsky, D., Dupraz, C., Madden, S., Viallefond, F., Vigroux, L., 1998, 
{\it A\&A}, 335, 53
\reference Bruzual, A.G., Charlot, S., 1993, {\it ApJ}, 405, 538
\reference Calzetti, D., Kinney, A.L., Storchi--Bergmann, T., 1994, 
{\it ApJ}, 429, 582
\reference Charbonnel, C., Meynet, G., Maeder, A., Schaerer, D., 1996,
{\it A\&AS}, 115, 339
\reference Cimatti, A., Andreani, P., Rottgering, H., Tilanus, R., 1998,
{\it Nature}, 1998a, 392
\reference Cimatti, A., Freudling, W., Rottgering, H., Ivison, R.J.,
Mazzei, P., 1998b, {\it A\&A}, 329, 399
\reference Devriendt, J.E.G., Sethi, S., Guiderdoni, B., Nath, B., 1998,
{\it MNRAS}, 298, 708
\reference Devriendt, J.E.G., Guiderdoni, B., Sadat, R., 1999, 
{\it A\&A}, 350, 381
\reference Devriendt, J.E.G., Guiderdoni, B., 1999, {\it submitted}
\reference Eales, S., Lilly, S., Gear, W., Dunne, L., Bond, J.R., Hammer, F., 
Le F\`evre, O., Crampton, D., 1999, {\it ApJ}, 515, 518
\reference Elbaz D., Aussel H., Cesarsky C.J., Desert F.X., Fadda D., 
           Franceschini A., Harwit, M., Puget J.L., Starck J.L., 1999,
           in {\it The Universe as seen by ISO}, P. Cox \& M.F. Kessler (eds),
           1998, UNESCO, Paris, ESA Special Publications series (SP-427)
\reference Fioc, M., Rocca--Volmerange, B., 1997, {\it A\&A},
326, 950
\reference Fixsen, D.J., Dwek, E., Mather, J.C., Bennett, C.L., Shafer, R.A.,
1998, {\it ApJ}, 508, 123 
\reference Flores, H., Hammer, F., Thuan, T.X., Cesarsky, C., D\'esert,
F.X., Omont, A., Lilly, S.J., Eales, S., Crampton, D., Le F\`evre., O., 1999, 
{\it ApJ}, 517, 148
\reference Guiderdoni, B., Bouchet, F.R., Puget, J.L., Lagache, G.,
Hivon, E., 1997, {\it Nature}, 390, 257 
\reference Guiderdoni, B., Hivon, E., Bouchet, F.R., Maffei, B., 1998, 
{\it MNRAS}, 295, 877
\reference Hauser, M.G., Arendt, R., Kelsall, T., Dwek, E., Odegard, N., 
Welland, J., Freundenreich, H., Reach, W., Silverberg, R., Modeley, S., 
Pei, Y., Lubin, P., Mather, J., Shafer, R., Smoot, G., Weiss, R., 
Wilkinson, D., Wright, E., 1998, {\it ApJ}, 508, 25
\reference Hughes, D., Dunlop, J.S., Rawlings, S., 1997, {\it MNRAS}, 289, 766
\reference Hughes, D., Serjeant, S., Dunlop, J., Rowan--Robinson, M.,
Blain, A., Mann, R.G., Ivison, R., Peacock, J., Efstathiou, A., Gear, W., 
Oliver, S., Lawrence, A., Longair, M., Goldschmidt, P., Jenness, T., 
1998, {\it Nature}, 394, 241
\reference Kawara, K., Sato, Y., Matsuhara, H., Taniguchi, Y., Okuda, H., 
Sofue, Y., Matsumoto, T., Wakamatsu, K., Karoji, H., Okamura, S., Chambers, 
K.C., Cowie, L.L., Joseph, R.D., Sanders, D.B., 1998, 
{\it A\&A}, 336, L9
\reference Kennicutt, R.G., 1998, in {\it Starbursts: Triggers, Nature and 
Evolution}, B. Guiderdoni \& A. Kembhavi (eds), EDP Sciences/Springer--Verlag
\reference Kurucz, R., 1992, IAU Symposium 149, 225
\reference Lagache, G., Abergel, A., Boulanger, F., D\'esert, F.X., 
Puget, J.L., 1999, {\it A\&A}, 344, 322
\reference Lilly, S.J., Eales, S.A., Gear, W.K.P., Hammer, F., Le F\`evre, O.,
Crampton, D., Bond, J.R., Dunne, L., 1999, {\it ApJ}, 518, 641
\reference Meurer, G.R., Heckman, T.M., Leitherer, C., Kinney, A. Robert, C., 
Garnett, D.R., 1995, {\it AJ}, 110, 2665
\reference Meurer, G.R., Heckman, T.M., Calzetti, D., 1999, 
{\it ApJ}, 521, 64
\reference Puget, J.L., Abergel, A., Bernard, J.P., Boulanger, F., 
Burton, W.B., D\'esert, F.X., Hartmann, D., 1996, {\it A\&A}, 
308, L5
\reference Puget, J.L., Lagache, G., Clements, D.L., Reach, W.T.,
Aussel, H., Bouchet, F.R., Cesarsky, C., D\'esert, F.X., Dole, H., Elbaz, D.,
Franceschini, A., Guiderdoni, B., Moorwood, A.F.M., 1999, 
{\it A\&A}, 345, 29
\reference Rigopoulou, D., Lawrence, A., Rowan--Robinson, M., 1996, 
{\it MNRAS}, 278, 1049
\reference Smail, I., Ivison, R.J., Blain, A.W., 1997, {\it ApJ}, 
490, L5
\reference Smail, I., Ivison, R.J., Blain, A.W., Kneib, J.P., 1998, 
{\it ApJ}, 507, L21
\reference Soifer, B.T., Neugebauer, G., Franx, M., Matthews, K., 
Illingworth, G.D., 1998, {\it ApJ}, 501, L171
\reference Steidel, C.C., Adelberger, K.L., Giavalisco, M., Dickinson, M.,
Pettini, M., 1999, {\it ApJ}, 519, 1
\reference Thuan, T.X., Sauvage, M., Madden, S., 1999, {\it ApJ}, 
516, 783
\end{references}
\end{document}